\documentclass[twocolumn,showpacs,preprintnumbers,amsmath,amssymb]{revtex4}

\usepackage{graphicx}
\usepackage{dcolumn}
\usepackage{bm}
\usepackage{amsmath,amssymb}

\begin{document}

\preprint{APS/123-QED}

\title{Mixing-Demixing transition in one-dimensional boson-fermion mixtures}

\author{Yousuke Takeuchi}
\affiliation{%
Department of Quantum Matter, ADSM, Hiroshima University,\\ 1-3-1 Kagamiyama, 
Higashi-Hiroshima 739-8530, JAPAN
}%

\author{Hiroyuki Mori}
\affiliation{
Department of Physics, Tokyo Metropolitan University,\\
Minamiohsawa 1-1, Hachioji-shi, Tokyo 192-0397, JAPAN\\
}%

\date{\today}

\begin{abstract}
Mixing-demixing transition in one-dimensional mixtures of fermions and bosons is numerically investigated by changing various parameters such as number densities of each component, fermion-boson interactions, boson-boson interactions, and hopping energies. In most cases we found clear evidences of the mixing-demixing transitions and identified the roles of each microscopic parameter in the transitions. Several phase diagrams were obtained.
\end{abstract}

\pacs{Valid PACS appear here}
\maketitle
\section{\label{intro}Introduction}
Recent studies of ultracold trapped atomic gases have led to the discovery of intriguing physical phenomena and have attracted more and more attentions to the atomic systems. The trapped atomic gases have some advantages as a physical system in that any combinations of bosons and fermions can be put together on a lattice, microscopic parameters are tunable in most cases, and the systems can be constructed in various geometries, e.g. one-dimensional chain or two-dimensional plane.

One dimensional systems were constructed with bosons \cite{Maretal}-\cite{Morietal} and with bosons and fermions \cite{Modetal}. These could be the first case that realized one-dimensional boson/boson-fermion systems, and triggered theoretical studies \cite{Aletal}-\cite{CazandHo}. 

At a commensurate filling with the lattice periodicity, the mixtures of spinless fermions and bosons in one-dimensional optical lattice were predicted to undergo  Mott transition at sufficiently large fermion-boson and boson-boson interactions \cite{Aletal, RotandBu}. 

In the boson-fermion mixtures at an incommensurate filling, on the other hand, the only possible instability is mixing-demixing transition. When the repulsive interactions between fermions and bosons are small, the two components are well mixed and the system is in the mixing state. When the interactions are large enough, each component stays away from the other and the system is in the demixing state. Cazalilla and Ho studied the mixing-demixing transition based on a Luttinger liquid formalism with a large difference in the number densities of the two components \cite{CazandHo}. They derived a criterion for the demixing of hardcore bosons and spinless fermions in the form, 
$\sqrt{\mathit{v}_f \mathit{v}_{Fb}} < |g_{fb}|/\hbar \pi$
where $g_{fb}$ is a fermion-boson interaction, $\mathit{v}_f$ is the Fermi velocity, and  $\mathit{v}_{Fb} = \hbar \pi \rho_{b}/M_{b}$ with $M_{b}$ being the atomic mass of boson and $\rho_{b}$ being the boson density. Also there is a rigorous proof that the system always remains in the mixing state as far as the fermion-boson interaction has the same magnitude as the boson-boson interaction \cite{AdiandEug}. For a slightly different but more generalized model, Batchelor \textit{et al.} found no demixing phase in the Bethe Anzatz solution \cite{batchelor}.

Under these circumstances we need more evidences for the existence of the mixing-demixing transition to draw clear phase diagrams. In this paper, we investigated numerically the mixing-demixing transitions of incommensurate fermion-boson mixtures to complete the phase diagrams, and identified the roles of various microscopic parameters in the transition.

The paper is organized as follows. In Chapt. 2 we will introduce the model Hamiltonian and a correlation function we used as an indicator to determine the phases. Chapter 3 is devoted to the description of calculation results and Chapt. 4 to the conclusion of the present study.
\section{\label{meth}Model}
We consider a mixture of $N_f$ spinless fermions and $N_b$ bosons on an $N$-site 
lattice. 
We employed Bose-Fermi Hubbard Hamiltonian to describe the interacting particles, which 
is defined by
 \begin{eqnarray}
\mathcal{H}&=&-\sum_i\sum_{\alpha=f,b}\left[ t_{\alpha}(a_{\alpha,i}^\dagger 
a_{\alpha,i+1}+h.c.)+\mu_{\alpha} n_{\alpha,i} \right] \nonumber\\
& &{}+\sum_i \left[ \frac{U_{bb}}{2} n_{b,i}(n_{b,i}-1)+U_{fb}n_{f,i}n_{b,i} \right],
\end{eqnarray}
where $a_{\alpha,i}^\dagger$ and $a_{\alpha,i}$ are respectively creation and annihilation 
operators for fermions ($\alpha =f$) or bosons ($\alpha =b$) on the $i$-th 
site, and $n_{\alpha,i}=a_{\alpha,i}^\dagger a_{\alpha,i}$.
Hopping energy and chemical potential are denoted respectively by $t_{\alpha}$ and $\mu_{\alpha}$.
$U_{bb}$ and $U_{fb}$ are on-site boson-boson and fermion-boson repulsive 
interactions respectively.

To observe the mixing-demixing transition, we define a correlation function 
\begin{equation}
C =\langle (n_{b,i-1}+n_{b,i+1})n_{b,i}\rangle - \langle 
(n_{f,i-1}+n_{f,i}+n_{f,i+1})n_{b,i}\rangle.\label{m1}
\end{equation}
The first term gives the correlation in neighboring bosons, and the second term 
expresses the correlation between bosons and fermions on the same and neighboring 
sites. In the mixing states fermions and bosons are spatially mixed and move throughout the 
system, while in the demixing states fermions and bosons reside separately and move in each 
restricted area. The correlation function $C$ changes drastically at the transition, reflecting these characteristics. It tends to be smaller, or negatively larger, in mixing states than in demixing states. The reason we omitted $\langle n_{b,i}n_{b,i}\rangle$ from the first term is that this 
sometimes dominates the correlation function when multiple bosons stay on the same 
site, mostly in the demixing states, and hides subtle behavior of the other terms. Many physical quantities change at the transition and we actually calculated other various physical quantities including the $i-j$ dependence of $\langle n_{b,i}n_{f,j}\rangle$. However we found the correlation function $C$ exhibited the most drastic change near the transition point and decided to use it as an indicator of the transition.
\section{\label{cal}Calculation and Result}
We performed Monte Carlo simulations with the world-line algorithm \cite{Hirsch, BatandSca} on a 
one-dimensional $N$-site lattice with periodic boundary conditions.
We changed in the simulations the number of fermions and bosons, $N_f$ and 
$N_b$ respectively, while fixing the total number of particles $N_f+N_b$ to 14. We used 
the hopping energy $t_b$ of bosons as an energy unit. Temperature was fixed to 
$T=0.08$ and the Trotter decomposition number $L$ to 100. The number of sites $N$ was set to 30.

In the following two subsections we will show the mixing-demixing transitions by changing various parameters with several choices of the difference in the number densities of the fermions and bosons, $\delta\rho =(N_f-N_b)/N$.
\subsection{Changing interactions}
\begin{figure}
\begin{tabular}{lclc}
\resizebox{35mm}{!}{\includegraphics{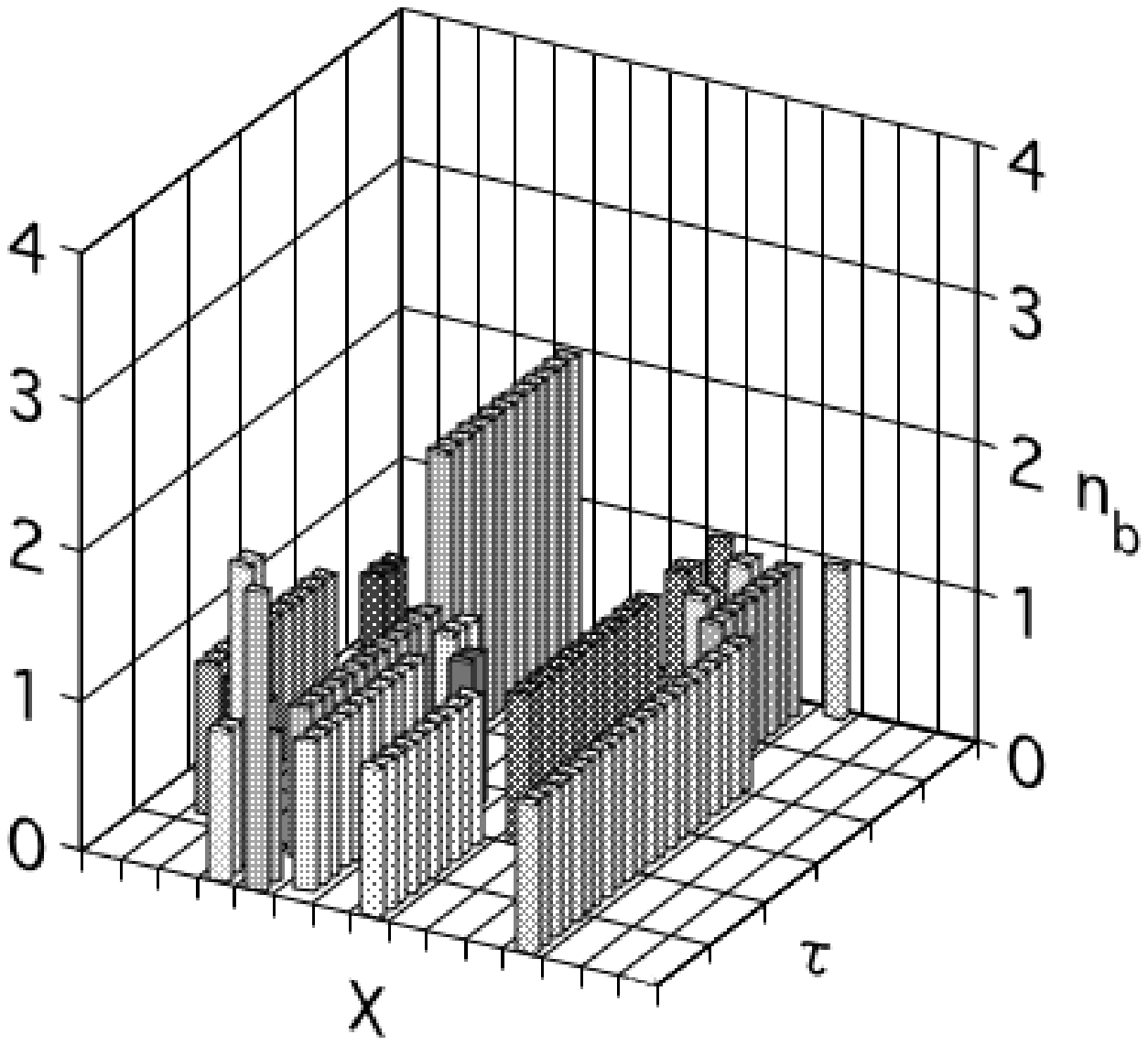}} & & 
\resizebox{35mm}{!}{\includegraphics{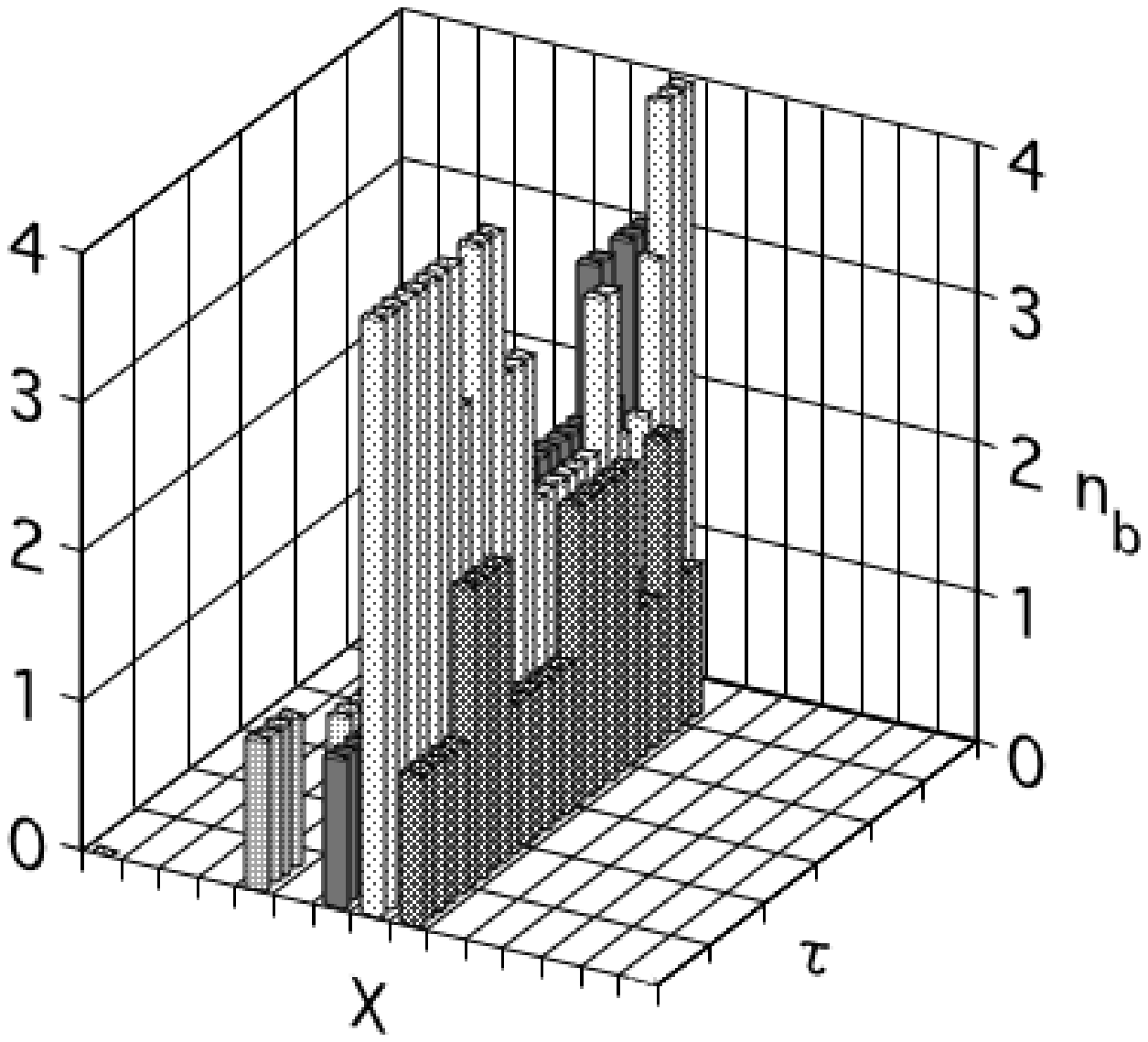}} &  \\
\resizebox{35mm}{!}{\includegraphics{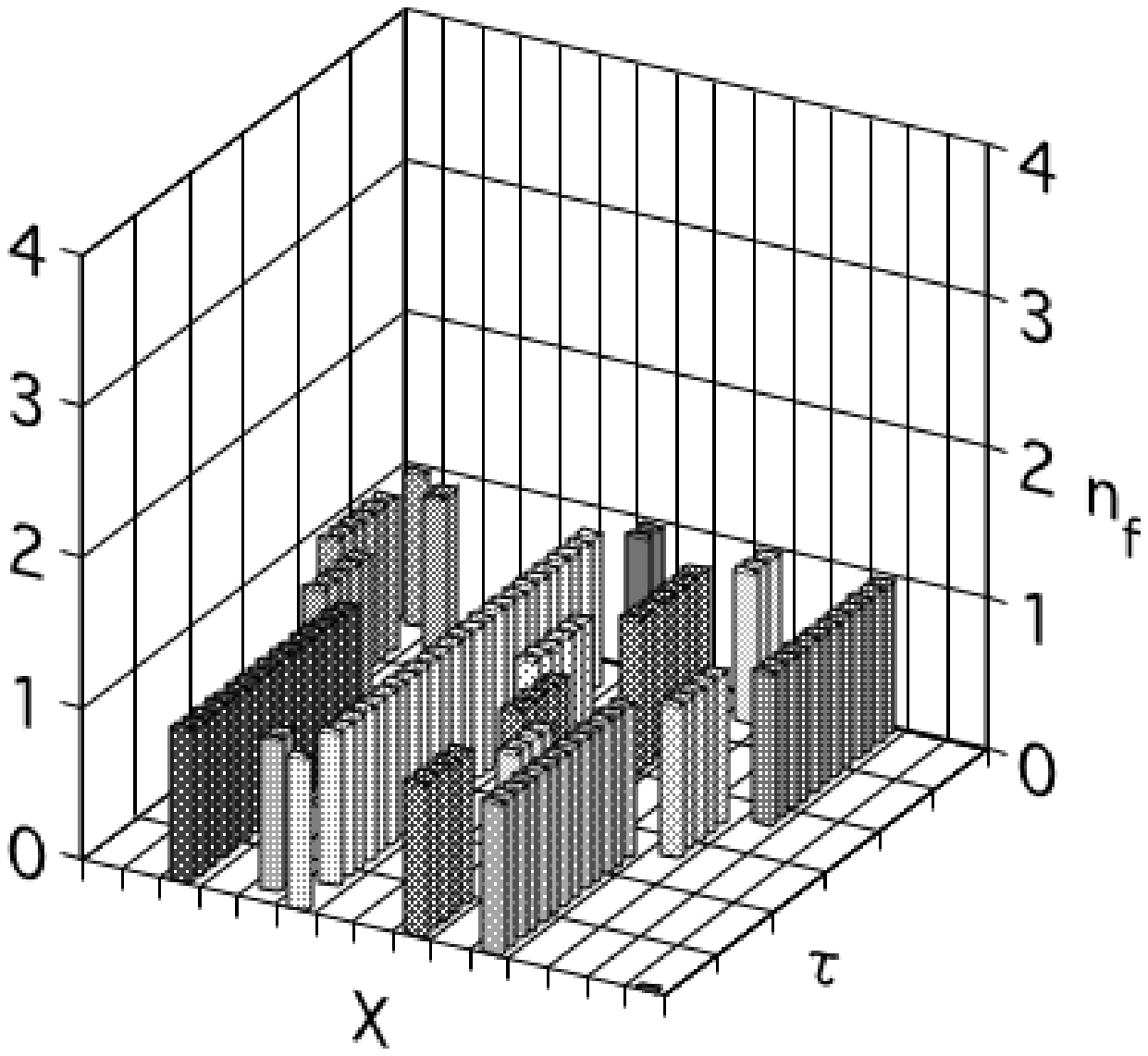}} 
&(a)&\resizebox{35mm}{!}{\includegraphics{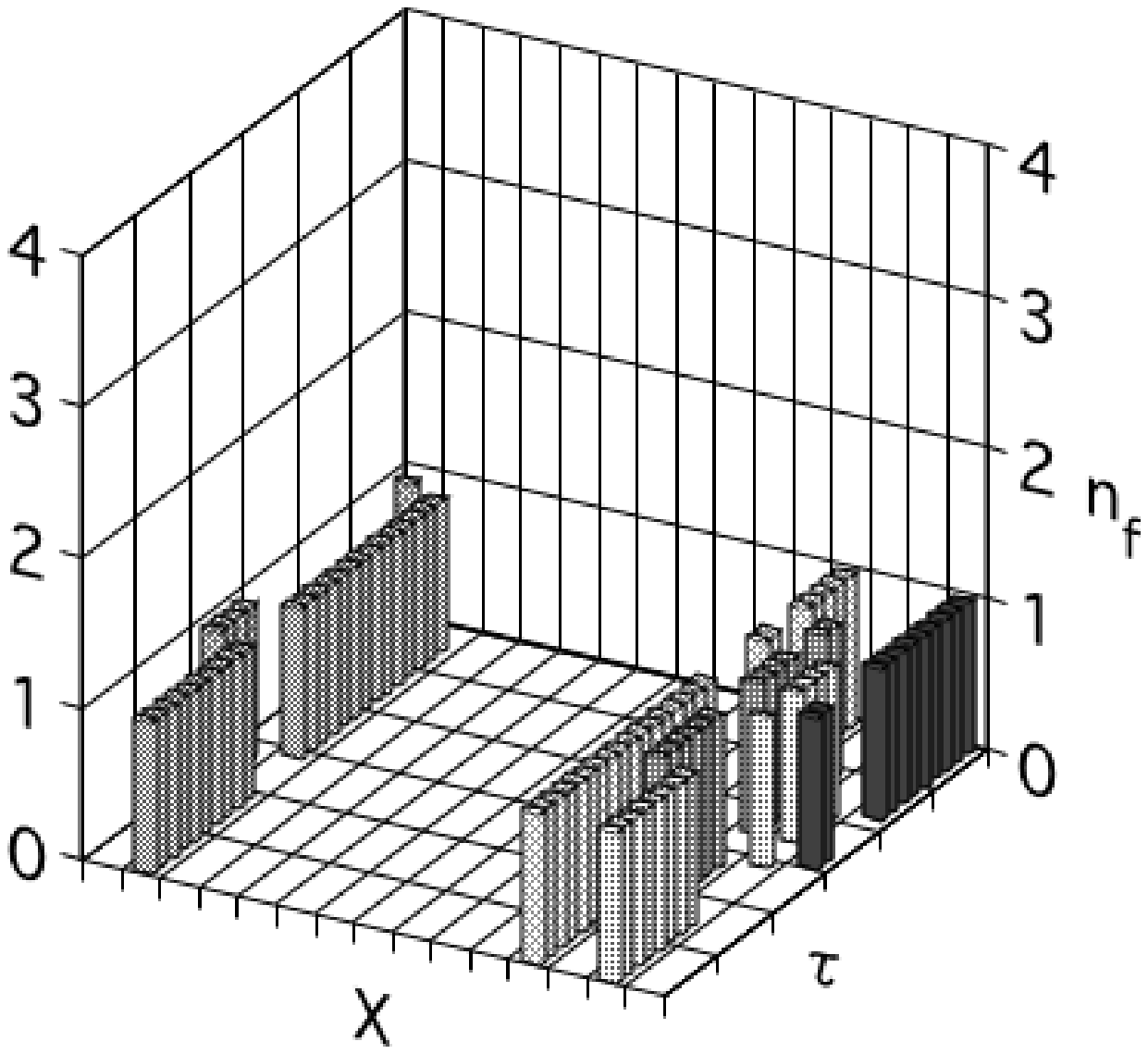}} & (b) 
\end{tabular}
\caption{\label{fig}The snapshot of fermions and bosons with $\delta\rho=0$, $t_f=1$ and $U_{bb}=0$. In (a) we set $U_{fb} =0.5$ and in (b) $U_{fb}=6$. The upper graphs in (a) and (b) show the snapshots of bosons and the lower graphs show those of fermions. The real space is indicated by X axis, $\tau$ denotes the imaginary time axis, and $n_b$ ($n_f$) axis presents the number of bosons (fermions) on each site. The variation of the darkness is just a guide for the eyes.}
\end{figure} 
At first we observed the mixing-demixing transitions by changing the interactions with the hopping energy ratio fixed to $t_f/t_b=1$. Figure 1 presents typical snapshots of the fermions and 
bosons on the (real space)-(imaginary time) plane with $U_{bb}=0$ and $\delta\rho =0$. As far as the interaction $U_{fb}$ between the fermions and bosons is 
small, both components are well mixed as in Fig. 1(a), while they are separated at a large $U_{fb}$ as in Fig. 1(b). Due to the absence of the boson-boson interaction, more than 
one boson can occupy the same site with no additional energy cost and hence the bosons can provide a sufficiently large space for the fermions. Next we observed the 
mixing-demixing transitions in more quantitative way. Figure 2 shows the correlation 
function $C$ as a function of the fermion-boson interaction $U_{fb}$ with various 
$\delta\rho$ and $U_{bb}$. $U_{bb}$ was set to 0 in Fig. 2(a), 0.4 in (b) and 1 in 
(c). The difference in the number densities of the fermions and bosons, $\delta\rho 
=(N_f-N_b)/N$, was chosen to be 0, 2/30, 4/30, 6/30, and 8/30. As we can see in the 
figure, there is a characteristic uprise in every curve, indicating the transitions from 
the mixing to the demixing states. We repeated the same calculations in different system sizes and found the uprising behavior became sharper consistently in the larger size. The transition curves were still broad to locate the transition point accurately, although we can obtain from the figure an evidence for the existence of the mixing-demixing transitions and determine the approximate location of the mixing and demixing phases in a parameter space. Finite size scaling should be done in future to identify the precise transition point.

Several features of Fig. 2 should be noted. Firstly the transition occurs 
at any $\delta\rho$, not just at relatively large $\delta\rho$ as indicated in \cite{CazandHo}. It occurs even when $\delta \rho=0$. Secondly the transition 
point seems to shift to larger $U_{fb}$ side as the interboson interaction 
$U_{bb}$ increases.
\begin{figure}
\begin{tabular}{cc}
\resizebox{55mm}{!}{\includegraphics{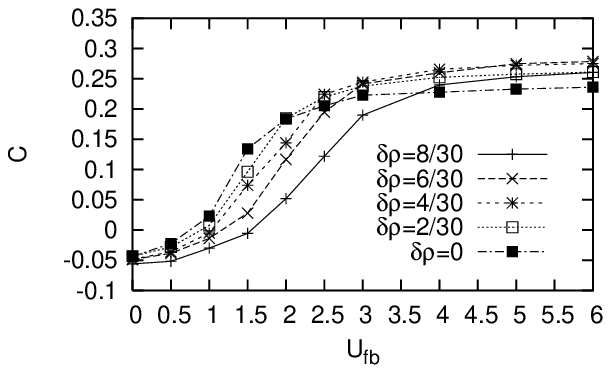}} & (a)\\ 
\resizebox{55mm}{!}{\includegraphics{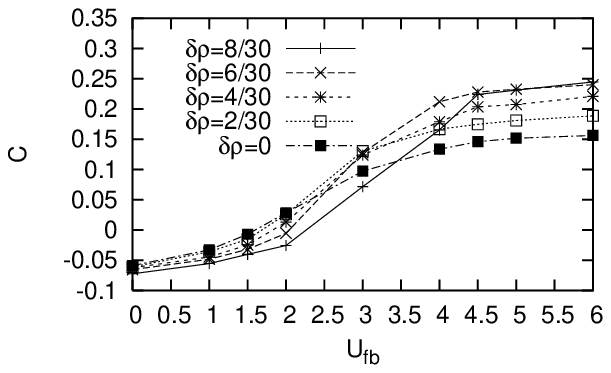}} & (b)\\ 
\resizebox{55mm}{!}{\includegraphics{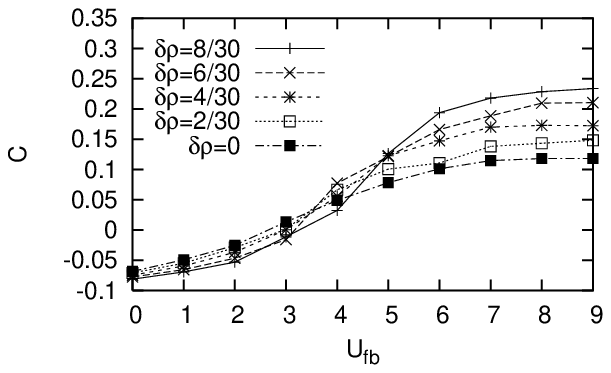}} & (c) 
\end{tabular}
\caption{\label{fig2} The correlation function $C $ as a function of $U_{fb}$ with 
$U_{bb}=0$ in (a), 0.4 in (b) and 1 in (c). $t_f$ is set to 1.}
\end{figure}

The second feature can be understood in the following way. In the demixing state, the bosons make a 
relatively small cluster by multiple occupation of sites as far as $U_{bb}$ is small, 
leaving a sufficiently large space for fermions to move around. Suppose we increase 
$U_{bb}$ in this state. It becomes hard for the bosons to stay in the small cluster. The 
boson cluster hence spreads to a certain extent to reduce the energy cost due to 
$U_{bb}$, which restricts the fermions to gain the kinetic energy. In other words, 
$U_{bb}$ helps mixing, while $U_{fb}$ helps demixing. Although we did not 
introduced interfermion interactions $U_{ff}$ in the model, it is obvious $U_{ff}$ would have the 
same effect as $U_{bb}$.

\begin{figure}
\begin{tabular}{cc}
\resizebox{40mm}{!}{\includegraphics{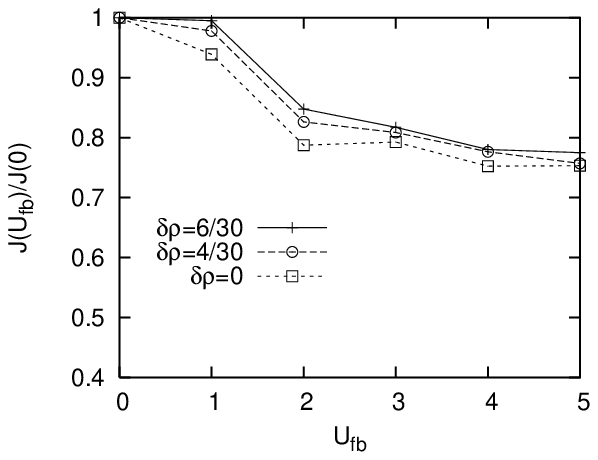}} & 
\resizebox{40mm}{!}{\includegraphics{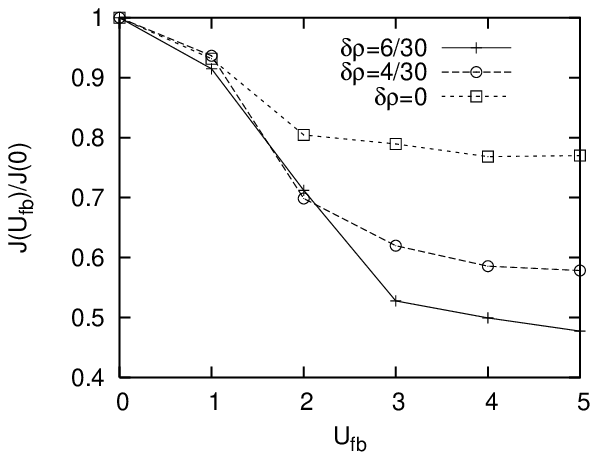}}\\
(a)&(b)
\end{tabular}
\caption{\label{fig9} Current-current correlation function $J$ of fermions (a) and bosons (b) as a function 
of $U_{fb}$ with $U_{bb}=0$ and $t_f =1$. The function $J$ is normalized by 
$J(U_{fb}=0)$.}
\end{figure}
We also observed the transitions by measuring another physical quantity, current-current correlation function $J_{\alpha =f,b}$ in the zero-frequency limit \cite{BatandSca, Batetal}:
\begin{equation}
J_{\alpha}=\lim_{\omega\to 0} \langle j_{\alpha}(\omega )j_{\alpha} (-\omega )\rangle ,
\end{equation}
where $j_{\alpha}$ presents the current density of the fermions when $\alpha =f$ and that of the bosons when $\alpha =b$. The correlation function $J$ of each kind of quantum particles is expected 
to be large in the mixing states where bosons and fermions extend over the system, and to be small in the demixing states where bosons and fermions are confined in their own restricted areas. 
Fig. ~\ref{fig9} shows the $J$ as a function of $U_{fb}$ with $t_f=1$ and 
$U_{bb}=0$. 
Every curve in the figure suggests the transition from mixing to demixing and the 
transition point is located somewhere in between $U_{fb}=1$ and 3, which is 
consistent to what we have seen in Fig. 2(a).

\begin{figure}
\begin{tabular}{cc}
\resizebox{50mm}{!}{\includegraphics{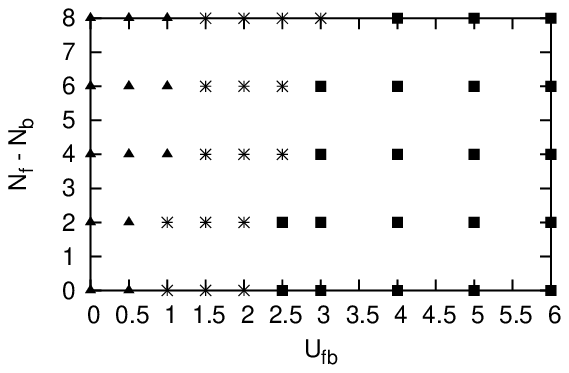}}&(a)\\ 
\resizebox{50mm}{!}{\includegraphics{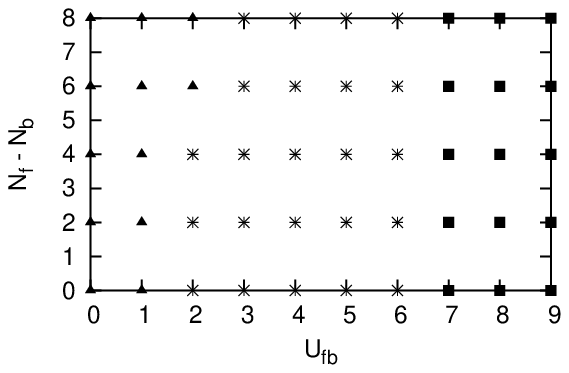}}&(b)\\ 
\end{tabular}
\caption{\label{fig6} Phase diagram in $N_f-N_b$ versus $U_{fb}$ with $U_{bb}=0$ (a) and $U_{bb} =1$ (b). $t_f$ is set to 1. "$\bigtriangleup$" and "$\Box$" present the 
mixing and the demixing states, respectively. "$\ast$" is marked for undetermined phase points.}
\end{figure}
Let us now draw the phase diagrams. Figure 4 shows 
the phase diagram in the $\delta\rho$-$U_{fb}$ plane. "$\bigtriangleup$" and 
"$\Box$" denote the mixing and demixing states respectively that can be determined 
clearly from the data in Figs. 1-3. As stated in the above, the transition curves are too broad to identify the precise location of the transition point and therefore we used the symbol "$\ast$" in the phase diagram to present a gray zone where we cannot determine whether the particles are 
mixed or demixed. However it is still clear that increasing $U_{bb}$ drives the transition area, presented by $\ast$, toward the larger $U_{fb}$ side. 
\subsection{Changing $t_f$}
\begin{figure}
\begin{tabular}{lclc}
\resizebox{35mm}{!}{\includegraphics{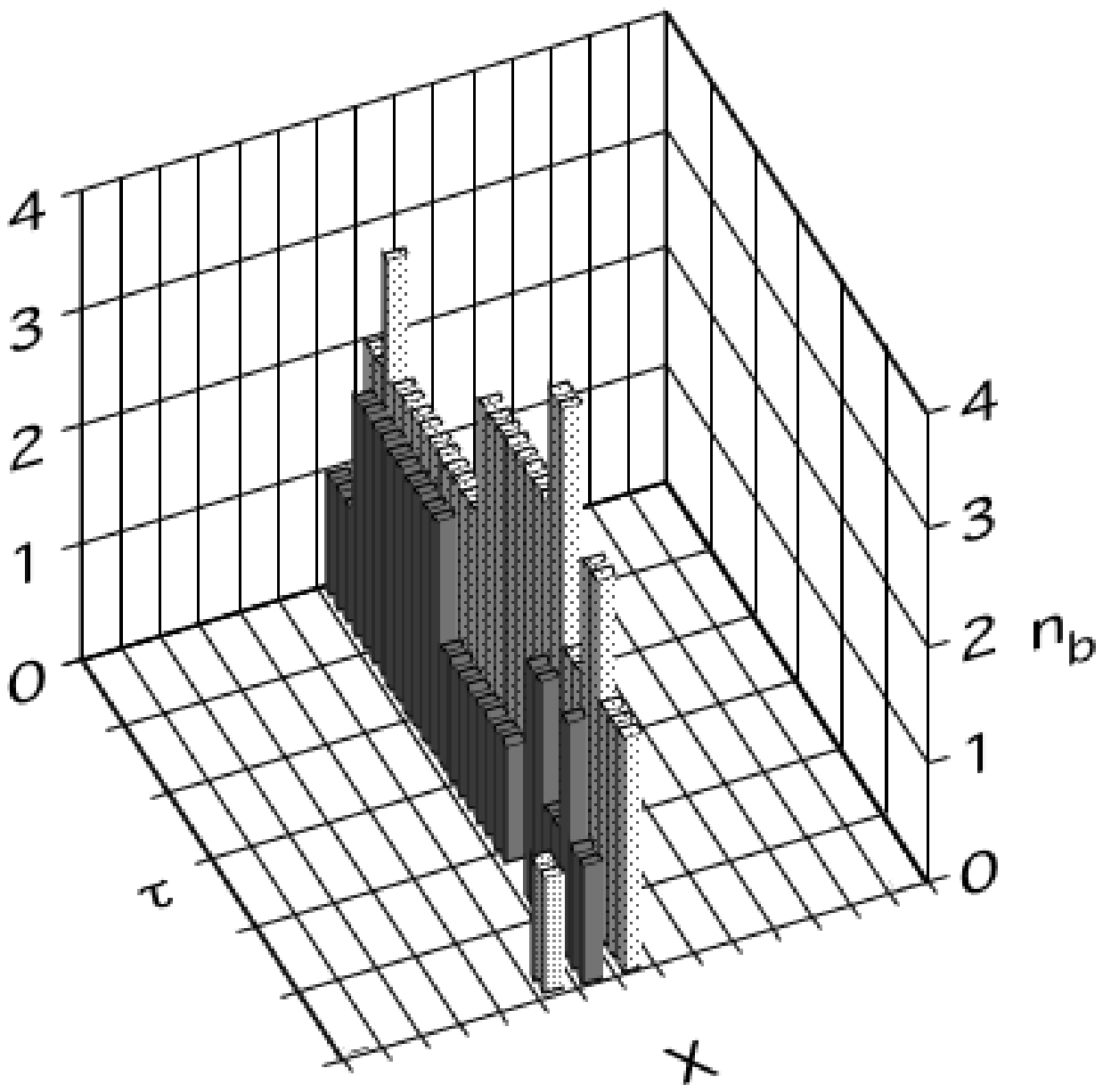}} & & \resizebox{35mm}{!}{\includegraphics{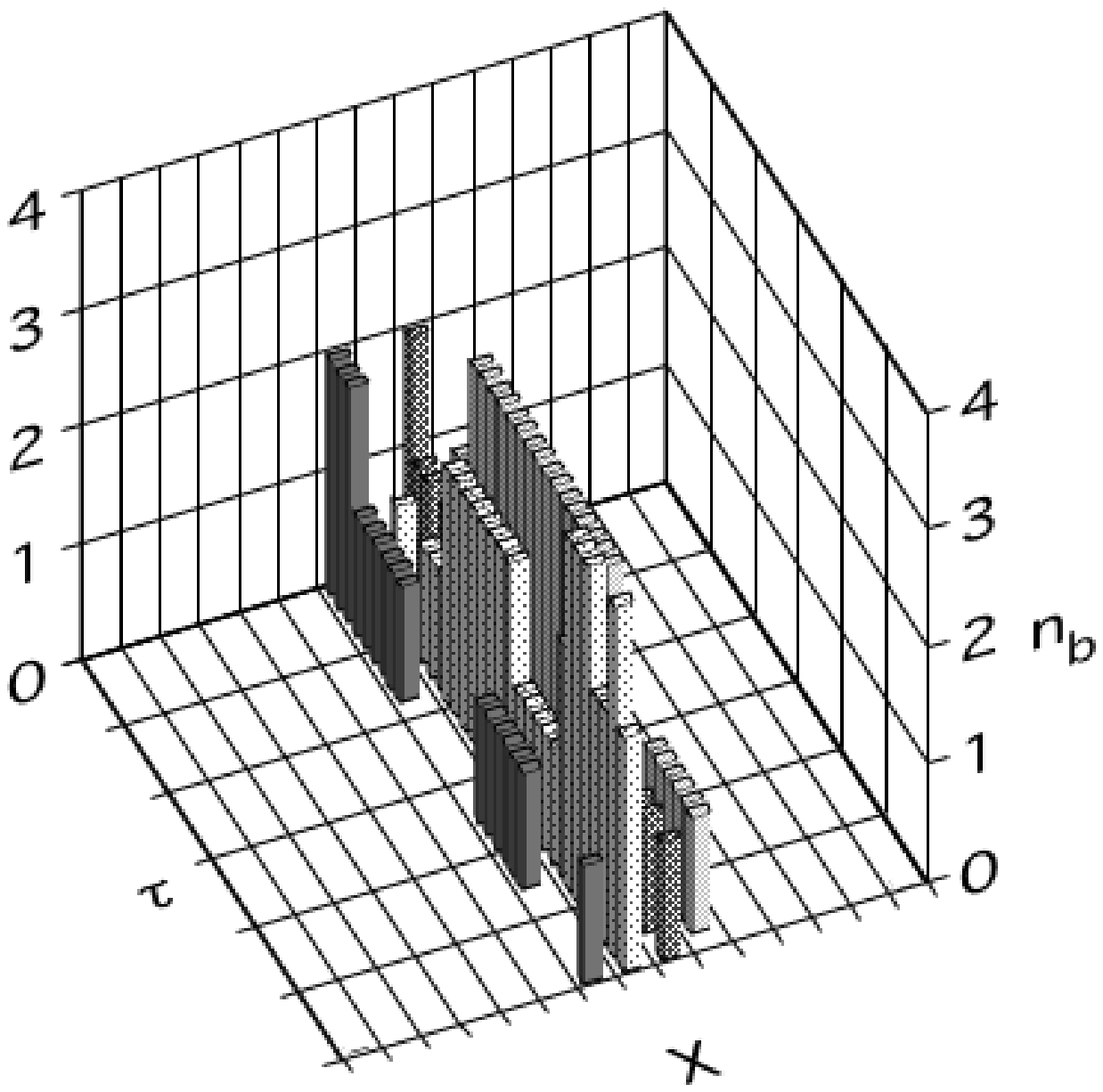}} &  \\
\resizebox{35mm}{!}{\includegraphics{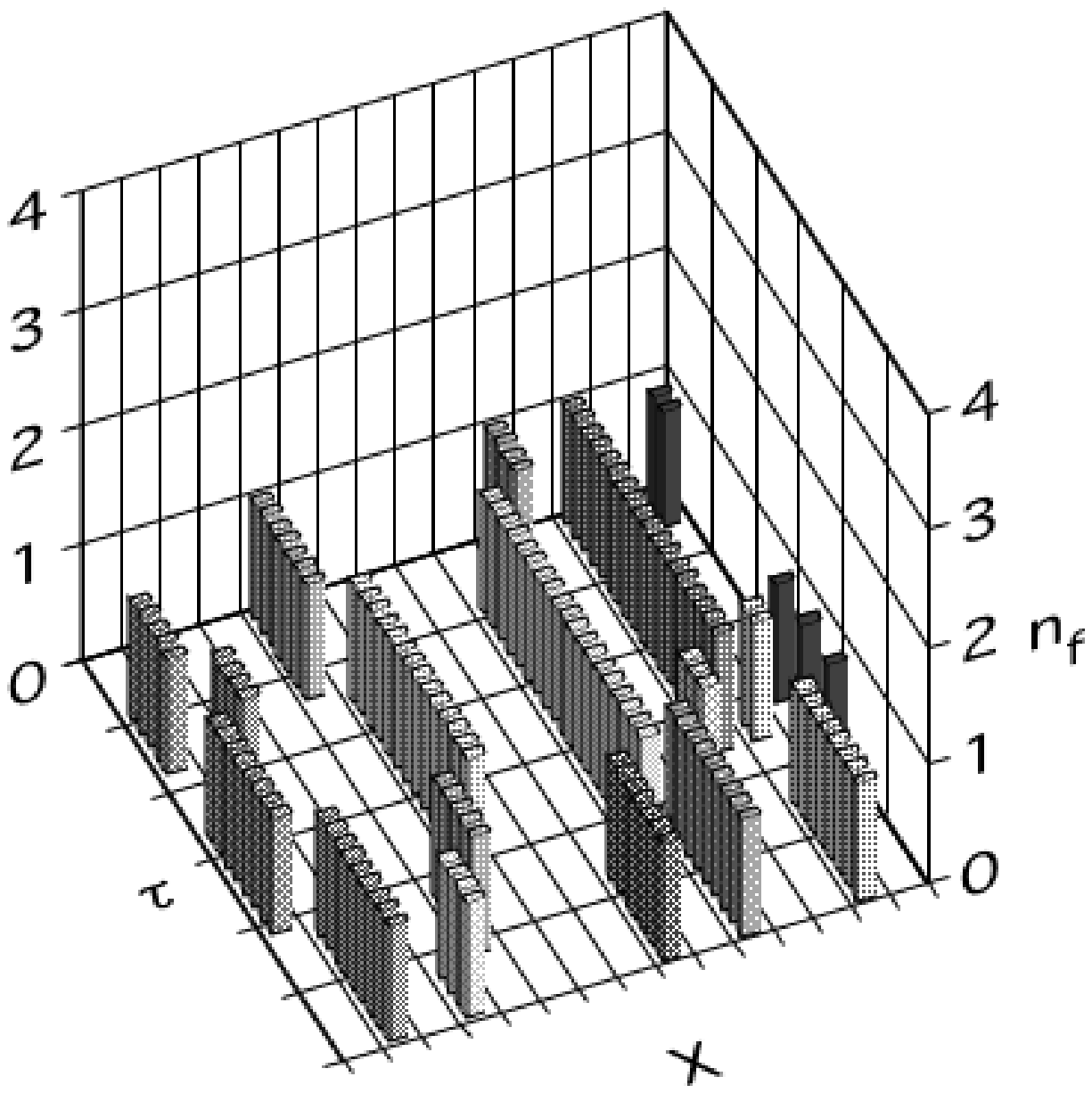}} &(a)&\resizebox{35mm}{!}{\includegraphics{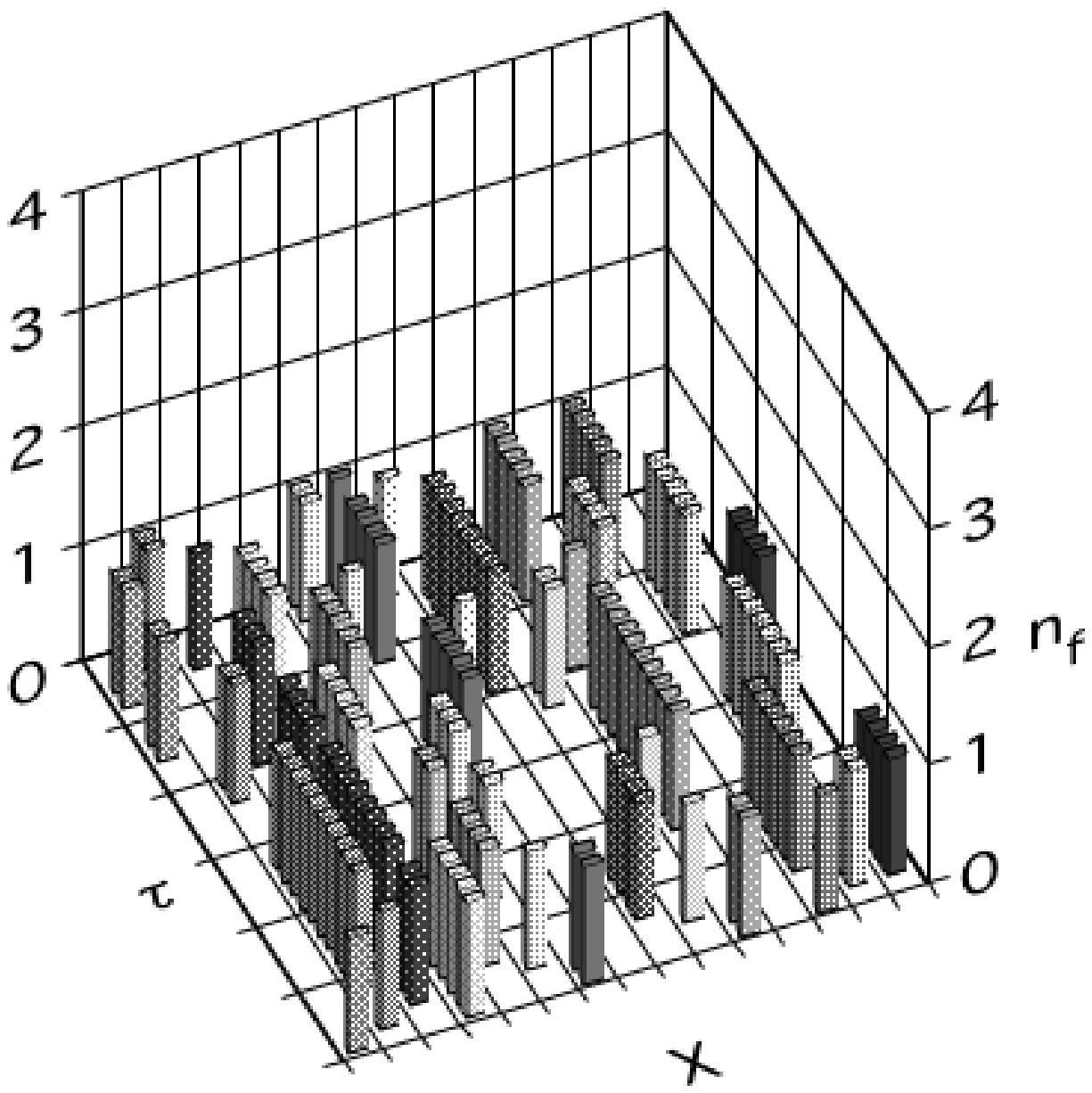}} & (b)
\end{tabular}
\caption{\label{fig11}The snapshot of fermions and bosons with $\delta\rho=6/30$, $U_{fb}=6$ and $U_{bb}=0$. In (a) we set $t_f =2$ and in (b) $t_f=7$. See the caption of Fig. 1 for other details of the figure.}
\end{figure}
We then observed the transitions by changing $t_f$.
Figure \ref{fig11} shows the snapshot of the bosons and fermions as in Fig. 1, although the hopping energy $t_f$ is increased here. By comparing (a) and (b), we see that the fermions are invading the boson area and making the system mixed as $t_f$ becomes larger.

\begin{figure}
\begin{tabular}{cc}
\resizebox{55mm}{!}{\includegraphics{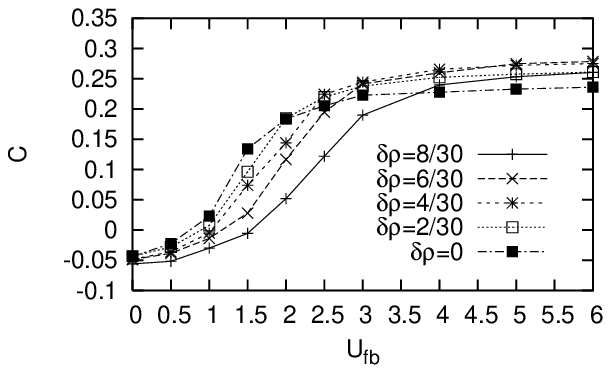}} & (a)\\ 
\resizebox{55mm}{!}{\includegraphics{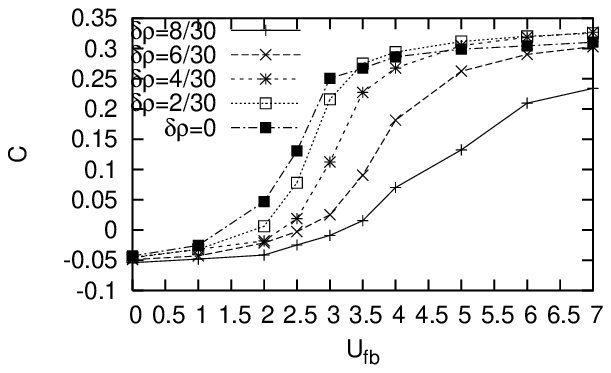}} & (b) 
\end{tabular}
\caption{\label{fig5} The correlation function $C $ as a function of $U_{fb}$ with 
$U_{bb} =0$. $t_f$ is 1 in (a) and 4 in (b). (a) is equivalent to Fig. 2(a).}
\end{figure}

Figure \ref{fig5} shows the correlation 
function $C$ versus $U_{fb}$ for two different values of $t_f$. Figure \ref{fig5} (a) is 
equivalent to Fig. 2(a) with the same parameters, and Fig. \ref{fig5} (b) shows the correlation 
function for larger $t_f$. As we see in the figure the transition points move to the larger 
$U_{fb}$ side as $t_f$ increases. In general, when hopping energy increases, 
particles need more space to gain the kinetic energy. This tendency is 
more drastic in fermions than in bosons as far as the on-site interboson interaction is 
finite. With the increase of $t_f$ in the demixing phase, the fermions try to invade the 
boson territory. Namely, $t_f$ helps mixing of the two components. For this reason the 
mixing-demixing transition occurs at a larger $U_{fb}$ when $t_f$ becomes large as in 
Fig.~\ref{fig5}.
\begin{figure}
\begin{tabular}{cc}
\resizebox{40mm}{!}{\includegraphics{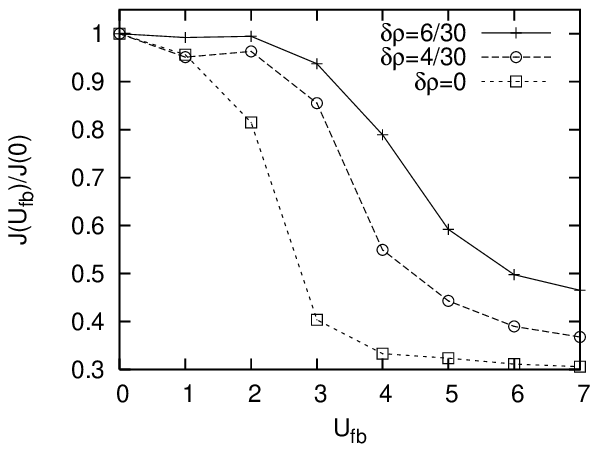}} & 
\resizebox{40mm}{!}{\includegraphics{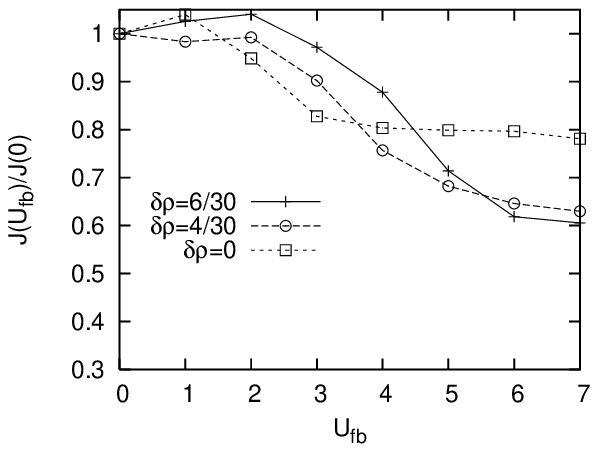}}\\
(a)&(b)
\end{tabular}
\caption{\label{fig10} Correlation function $J$ of the fermions (a) and the bosons (b) as a function 
of $U_{fb}$ with $U_{bb}=0$ and $t_f =4$. The function $J$ is normalized by 
$J(U_{fb}=0)$.}
\end{figure}
\begin{figure}
\begin{tabular}{c}
\resizebox{50mm}{!}{\includegraphics{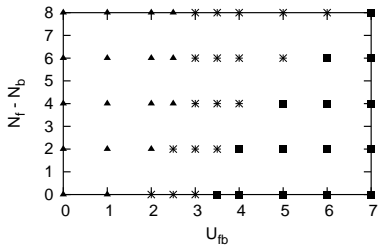}}
\end{tabular}
\caption{\label{fig7} Phase diagram in $N_f-N_b$ versus $U_{fb}$ with
$U_{bb} =0$ and $t_f =4$. "$\bigtriangleup$" and "$\Box$" present the mixing and the demixing 
states, respectively. "$\ast$" is marked for undetermined phase points.}
\end{figure}

The same behavior can be found in Fig.~\ref{fig10} where the correlation function $J$ of the fermions (a) and the
bosons (b) is calculated. Figure \ref{fig10} should be compared with Fig. 3. As $t_f$ increases 
the transition points shift to the larger $U_{fb}$ side.

We also drew the phase diagram in Fig. ~\ref{fig7} with a larger $t_f$ than used in Fig. 4 (a). In 
comparison of Fig.~\ref{fig7} with Fig. 4(a) we see the transition area, represented by $\ast$, 
clearly shifts to the larger $U_{fb}$ side.

\section{\label{con}Conclusion} 
\begin{figure}
\begin{tabular}{c}
\resizebox{50mm}{!}{\includegraphics{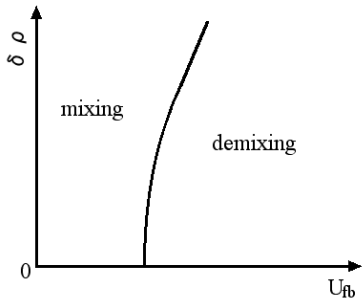}}
\end{tabular}
\caption{\label{phase} Schematic phase diagram of the mixing and the demixing phases.}
\end{figure}

We studied the mixing-demixing transitions of fermion-boson mixtures in one 
dimension by changing various parameters.
Monte Carlo calculation showed the mixing-demixing transition occurred at any 
$\delta\rho$, the difference in the number densities of the two components, as the 
repulsive interactions between the fermions and bosons increase. The schematic phase 
diagram is given in Fig.~\ref{phase}. 
Also we confirmed the role of each parameter. The fermion-boson interaction 
$U_{fb}$ helps demixing, while the boson-boson interaction $U_{bb}$ and the 
hopping energy $t_f$ help mixing. The phase boundary of Fig.~\ref{phase} therefore shifts to the 
larger $U_{fb}$ side as $U_{bb}$ or $t_f$ increases.

In our simulations we could not obtain reliable results at large interactions because of the restriction on the computational time. We performed the simulations for fermions and hardcore bosons, i.e. with $U_{bb}=\infty$, and found no demixing phase as far as $U_{fb}\le 20$. Although we have no numerical evidence on the absence of the demixing phase at $U_{fb}> 20$, we believe the fermions and the hardcore bosons would not be demixed at any $U_{fb}$ for the following reason. Suppose we have demixed fermions and hardcore bosons with a finite $U_{fb}$. In the vicinity of the boundary, the fermions and bosons move around, interacting with each other. The fermions (bosons) can only move to empty sites or sites occupied by the bosons (fermions), which means that it is energetically favored for each component of the particles to invade the area of the other component rather than to stay in its own area. In other words the demixing state of the fermions and the hardcore bosons would be unstable against mixing. In fact there is a rigorous proof that a mixture of fermions and bosons remains in the mixing state when $U_{fb}=U_{bb}$ \cite{AdiandEug}, indicating finite $U_{fb}$ cannot demix the mixture of fermions and hardcore bosons.

\begin{acknowledgments}
The authors would like to thank Prof. Jo and Prof. Oguchi for their continuous support.
\end{acknowledgments}


\end{document}